Force Generation by Myosin II Filaments in Compliant Networks
Samantha Stam[1,2], Jon Alberts[3], Margaret L. Gardel[2,4], Edwin Munro[2,5]

[1]Biophysical Sciences Graduate Program, [2]Institute for Biophysical Dynamics,
[3]University of Washington, Friday Harbor, Washington
[4]Physics Department & James Franck Institute,
[5]Department of Molecular Genetics and Cell Biology
University of Chicago, Chicago, IL 60637

Myosin II isoforms with varying mechanochemistry and filament size interact with filamentous actin (F-actin) networks to generate contractile forces in muscle and non-muscle cells. How their properties control force production in environments with varying stiffness remains poorly understood.  Here, we incorporated literature values for biophysical properties of myosin II isoforms into a cross-bridge model. Ensembles of motors produced similar actin gliding speeds and force-velocity curves expected from previous experiments. Their force output on an external elastic load was regulated by two timescales—that of their attachment to F-actin, which varied sharply with the ensemble size, motor duty ratio, and external load, and that of force build up, which followed a scaling relationship with ensemble stall force, gliding speed, and load stiffness. While such regulation did not require force-dependent kinetics, the myosin catch bond produced positive feedback between attachment time and force to trigger switch-like transitions from transient attachments generating small forces to high force-generating runs at threshold parameter values. Using appropriate parameters for skeletal muscle myosin, non-muscle myosin IIB, and non-muscle myosin IIA revealed three distinct regimes of behavior respectively: (1) large assemblies of fast, low-duty ratio motors rapidly build stable forces over a large range of environmental stiffness, (2) ensembles of slow, high-duty ratio motors serve as high-affinity cross-links with force build-up times that exceed physiological timescales, and (3) small assemblies of low-duty ratio motors operating at intermediate speeds are poised to respond sharply to changes in mechanical context—at low forces or stiffness, they serve as low affinity cross-links but they can transition to effective force production via the positive feedback mechanism described above. Together, these results reveal how myosin isoform properties may be tuned to produce force and respond to mechanical cues in their environment.

**Introduction**

Actomyosin contractility involves interactions of myosin II motors with actin filament (F-actin) arrays and powers a wide range of physiological processes including muscle contraction (1, 2), cell migration (3, 4), cell division (5, 6) and tissue morphogenesis (7, 8). These diverse contractile functions are mediated by functionally distinct myosin II isoforms operating within actin arrays that range from highly ordered muscle sarcomeres to more loosely organized contractile bundles to highly disordered networks. Contractile forces generated by myosin II are highly sensitive to mechanical context. This mechanosensitivity has been best studied in skeletal muscle, but may also allow non-muscle cells to sense and respond to mechanical signals such as external force or stiffness (9-11). However, we have yet to achieve a quantitative understanding of how myosin force generation depends on the interplay of motor properties and cellular mechanics.

All myosin II motors operate within larger bipolar ensembles known as myosin filaments, which vary in size from a few dozen heads for mini-filaments of non-muscle myosin II to hundreds of heads for the thick filaments of skeletal muscle myosin (12-16). Likewise, all myosin motors share a conserved mechanochemical cycle in which the energy of ATP hydrolysis is coupled to motor filament binding and the force generating powerstroke, but the rates of individual steps in this cycle vary widely across isoforms (17-22), leading to large differences in duty ratio (20-23) and unloaded F-actin gliding speed (23-25). Finally, a key feature shared by all myosin II isoforms is that the lifetime of the actin-bound state, and thus duty ratio, increases with opposing loads and decreases with assisting loads (26-28). Previous models of force-production by skeletal muscle myosin suggest that force-dependent release can enhance both tension and maximum shortening speed during contraction (29, 30). How force-dependent release affects the rate and magnitude of tension buildup by other myosin II isoforms in other cellular contexts remains poorly understood. Thus, a general challenge is to understand how isoform-specific tunings of force-dependent kinetics and filament size shape the rate, magnitude and mechanosensitivity of force production by ensembles of myosin II motors against dynamic and compliant actin arrays in living cells.

The swinging cross-bridge model for myosin II has played a key role in connecting the molecular properties of single motors to the macroscopic dynamics of contractile force production (31). The cross-bridge model has been used mainly in the context of skeletal muscle contraction where the large number of motors and sarcomeric organization facilitate simpler coarse-grained analyses (29, 32-37). More recently, cross-bridge models have been used to explore the dynamics of force production and filament translocation in non-sarcomeric contexts (30, 38, 39). However, these models have yet to be used in a more systematic exploration of how force production varies with isoform-specific motor properties, filament size, and substrate (i.e. F-actin network) compliance.

Here, we used computer simulations to explore how of the interplay of motor properties and environmental stiffness shapes the rate, stability and mechanosensitivity of force generation. We surveyed a range of motor duty ratios, filament sizes and gliding speeds that spanned measured values for different myosin isoforms and a range of environmental stiffness. We found that myosin motors build force more rapidly on stiffer substrates regardless of force-dependent kinetics. In general, both stability and magnitude of force generation increased with duty ratio and filament size. However, the interplay of force-dependent myosin release and external resistance sharpened this effect resulting in switch-like activation of force production for small changes in myosin filament size, duty ratio and environmental stiffness. Further analysis predicted the location of this transition with varying stiffness for motors with properties representing varying myosin II isoforms and revealed distinct regimes. Large ensembles of low duty ratio motors such as those found in skeletal muscle processively built force on elastic loads with a wide range of stiffness. Parameters more similar to myosin IIB motors also produced highly processive attachments, but the slow motor speed produced force build-up times that were prohibitively slow on relevant cellular timescales. Finally, small motor clusters representative of myosin IIA were poised to respond to the stiffness range we explored in the switch-like manner described above and transitioned from short attachments to processive force generation. These results therefore have important implications for understanding the regulation of contractile tension in different cellular contexts.

## METHODS
### Model Description

We extended the swinging cross-bridge model to ensembles of myosin II motors in a way that can be readily tuned to capture variation in mechanochemistry and filament size across different isoforms of myosin II. All myosin II motors share a basic mechanochemical cycle (40), with the following steps: (i) hydrolysis of ATP places the motor into a "primed" (Myo.ADP.Pi) state with high affinity for F-actin; (ii) Binding to F-actin is followed by Pi release and an internal conformation change (the "powerstroke") that converts the stored energy of ATP hydrolysis into force applied to the actin filament; (iii) ADP release followed by (iv) ATP rebinding and filament unbinding to complete the cycle.

We considered a two-state model in which the myosin cross-bridge switches between an unbound/weakly bound state and a tightly bound post-powerstroke state (Figure 1A). We represented the cross-bridge as an elastic element tethered at position $x_0$ to a rigid substratum representing the backbone of the myosin filament or the surface of a glass coverslip (Figure 1B). The cross-bridge bound actin in a pre-strained (post-powerstroke) state at $x = x_0 + d_{step}$ and exerted a force $F(x) = k_{x\text{-}bridge}(x-x_0)$ before unbinding. As in (30), we assumed the transition from unbound to bound state is first order with rate constant $k_{on}$ (Figure 1A); this precluded the possibility that the free energy cost of the powerstroke transition could become limiting under certain conditions (29, 38, 39).

Likewise, we subsumed ADP release, binding of ATP and release of the myosin head from F-actin into a single transition from bound to unbound states with rate $k_{off}$ (Figure 1A).

**Table 1: Parameter Values**

| Name | Description | Value | Reference |
|---|---|---|---|
| $N_{heads}$ | number of heads | variable; 2-1000 | see Table 2 |
| $k_{on}$ | binding rate | variable; 1-10 s$^{-1}$ | see Table 2 |
| $k_{off}(0)$ | unloaded unbinding rate | variable; 0.35-500 s$^{-1}$ | see Table 2 |
| $\alpha_{catch}$ | See Eqn (1) | 0.92 | (27) |
| $\alpha_{slip}$ | See Eqn (1) | 0.08 | (27) |
| $x_{catch}$ | See Eqn (1) | 2.5 nm | (27) |
| $x_{slip}$ | See Eqn (1) | 0.4 nm | (27) |
| $k_{off}(F)$ | force dependent unbinding rate | See Eqn (1) | (27) |
| $d_{step}$ | step size | 5.5 nm | (26) |
| $k_{x\text{-}bridge}$ | cross-bridge stiffness | 0.7 pN/nm | (26) |

Consistent with in vitro studies, we assumed that ATP was sufficiently abundant that binding and myosin release were fast relative to ADP release such that ADP release was the rate-limiting step for myosin detachment (17, 18, 20, 21).

All myosin II isoforms exhibit "catch-bond" behavior in which forces that oppose the motor (resisting loads) reduce the rate of motor detachment from F-actin, while assisting loads increase motor detachment (26-28). Above a critical force the bond behaves like a traditional slip bond (27). To represent this behavior, we used the force-dependent form of $k_{off}$ determined experimentally for skeletal muscle myosin II by Guo and Guilford (2006)

$$k_{off}(F) = k_{off}(0)\left[\alpha_{catch}\exp\left(-Fx_{catch}/k_BT\right) + \alpha_{slip}\exp\left(Fx_{slip}/k_BT\right)\right] \quad (1)$$

where the force *F* is positive for a resisting load, $k_B$ is Boltzmann's constant, *T* is temperature, $x_{catch}$ and $x_{slip}$ are characteristic bond lengths, and $\alpha_{catch}$ and $\alpha_{slip}$ are prefactors controlling the weights of the catch and slip components (Table 1). The unloaded detachment rate $k_{off}(0)$ can be tuned to model variation in detachment rates and duty ratios across different isoforms (see below).

**Simulations**

We considered a linear ensemble of myosin crossbridges attached at 5 nm intervals to a rigid substrate (Figure 1B), that bind and exert force upon an actin filament that is tethered at its barbed end by an elastic spring with stiffness *k* to reflect stiffness of the surrounding network. Binding sites for Myosin II were arrayed at 2.7 nm intervals. We sampled binding and unbinding rates stochastically to determine transitions between bound and unbound states (41). Between transitions, we computed instantaneous filament velocity by solving numerically the following equation of motion for the F-actin:

$$m\ddot{x} = 0 = -\gamma\dot{x} - F_{myo} + F_{ext} \tag{2}$$

where $F_{myo}$ is the total force exerted by the myosin crossbridges, $F_{ext}$ is the external force on the filament, and $\gamma$ is a drag coefficient chosen to approximate the hydrodynamic drag on a 10 ☐m long F-actin in cytoplasm (assuming a cytoplasmic viscosity 100 fold higher than the viscosity of water). This method is inefficient relative to approaches that assume instantaneous mechanical relaxation between binding/unbinding events (30, 38, 39). We chose to use it here because it extends naturally to simulations of larger motor/filament ensembles and because computational time was not rate-limiting for our analysis.

**Table 2: Tunable parameter values used to represent myosin isoforms in Fig. 1**

| Parameter | Isoform | Value | Reference |
|---|---|---|---|
| $k_{off}(0)$ | skeletal | 500 s$^{-1}$ | (19) |
|  | smooth | 22 s$^{-1}$ | (17, 18) |
|  | nonmuscle IIA | 1.71 s$^{-1}$ | (20) |
|  | nonmuscle IIB | 0.35 s$^{-1}$ | (21) |
| $k_{on}$ | skeletal | 10 s$^{-1}$ | (42) |
|  | smooth | 1 s$^{-1}$ | (18) |
|  | nonmuscle IIA | 0.2 s$^{-1}$ | (20) |
|  | nonmuscle IIB | 0.2 s$^{-1}$ | (21) |
| $N_{heads}$ | skeletal | 500 | (14) |
|  | smooth | 300 | (15, 16) |
|  | nonmuscle IIA | 50 | (12) |
|  | nonmuscle IIB | 50 | (12) |

**RESULTS**

**Benchmarking motor performance for different myosin II isoforms**

We first examined how isoform-specific variation in motor properties and filament size influence force buildup and maintenance. As a first step, we tested the ability of our model to reproduce isoform-specific variation in the gliding filament assay--a standard measure of ensemble motor performance. We assigned isoform-specific values for the attachment rate $k_{on}$ and the unloaded detachment rate $k_{off}(0)$ based on *in vitro* studies (see Table 2). For each set of parameters, we measured the unloaded gliding velocity ($V_{max}$) as a function of the number of myosin heads ($N_{heads}$). In all four cases, $V_{max}$ increased monotonically with $N_{heads}$ and saturated at high values (Fig. 1C). The maximal gliding velocities were in good agreement with those observed experimentally for these four isoforms (23-25). The saturation of $V_{max}$ with increasing $N_{heads}$ was consistent with experimental gliding filament assays (23, 43) and previous models (30, 38) and reflected the transition to a "detachment-limited" regime in which newly attached motors face increasing opposition from previously attached cross-bridges that become negatively strained before detachment (data not shown). Notably, the isoform-specific curves collapsed when we scaled the velocity by $V_{saturation}$,, the value observed at high (saturating) values of $N_{heads}$, and the x-axis by the unloaded duty ratio $dr(0)$, where $dr(0) = k_{on}/(k_{on} + k_{off}(0))$ (Fig. 1D).

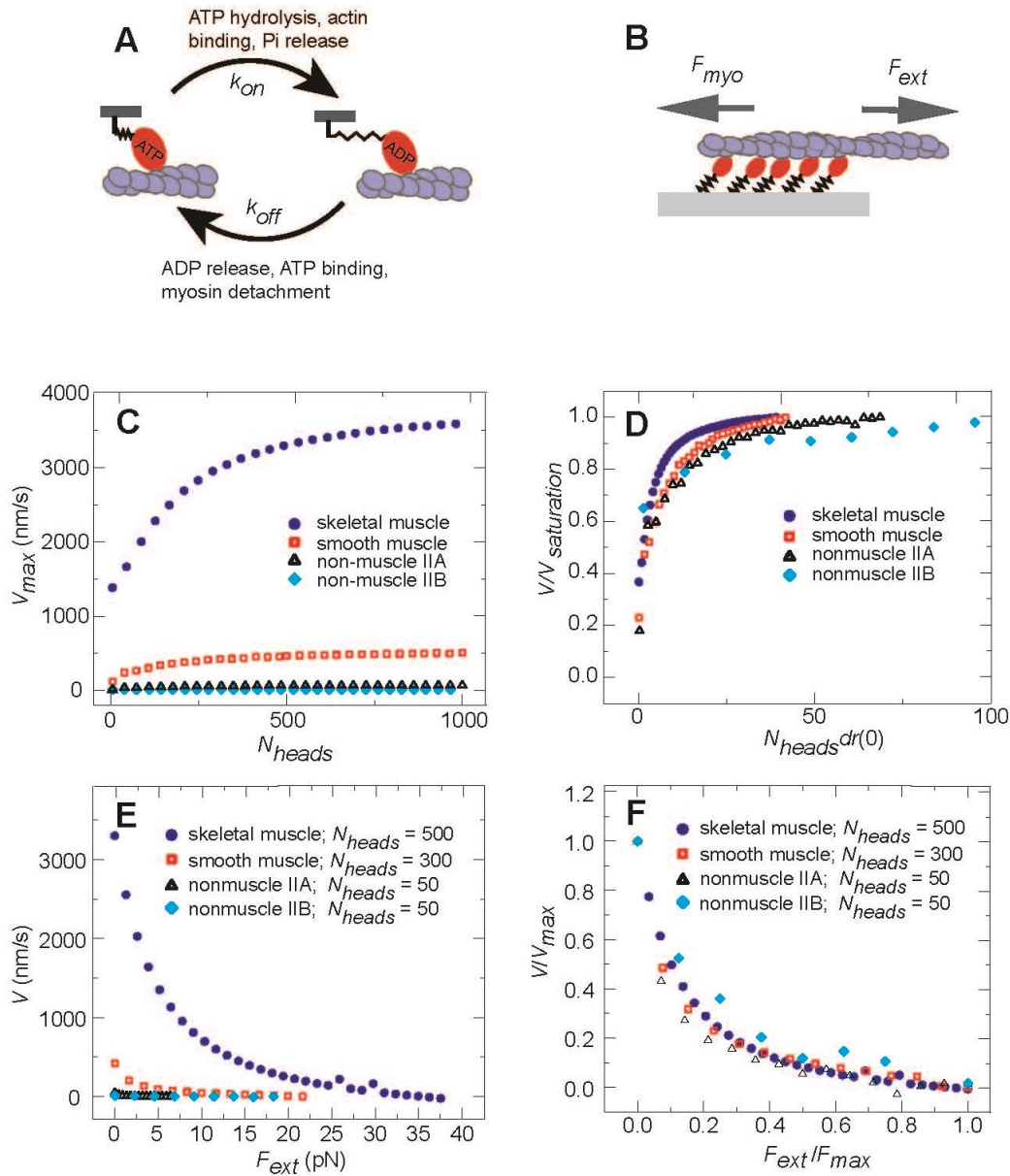

Figure 1: Model simulations reproduce the expected gliding velocity and force-velocity curves for motor clusters. (A) Description of simplified two-step mechanochemical cycle: myosins strongly attached to actin filaments at a rate $k_{on}$ and detached with rate $k_{off}$. (B) Simulation setup: motor heads were attached to a fixed surface. The heads pulled against an external force $F_{ext}$. (C) The mean gliding velocity of an unloaded actin filament ($V_{max}$) as a function of $N_{heads}$. Only the velocity while at least one myosin head was attached was considered. (D) Scaling the vertical axis and the horizontal axis of (C) by the plateau velocity at large $N_{heads}$ and $dr(0)$ respectively resulted in collapse of the data. (E) Mean gliding velocity with varying $F_{ext}$. (F) Collapse of curves from (E) when the vertical and horizontal axes were scaled by their respective intercepts $V_{max}$ and $F_{max}$. In all panels, the default parameters from Table 1 and isoform-specific parameters from Table 2 were used.

Next, we considered how the gliding velocity varied with externally applied force, $F_{ext}$, for isoform-specific values of $k_{on}$, $k_{off}(0)$ and $N_{heads}$ appropriate for different isoforms (Table 2). For each parameter set, our simulations reproduced the concave force-velocity relationship observed experimentally (43, 44) and in previous models (29, 32-37). When we scaled velocity and force by $V_{max}$ and the ensemble stall force $F_{max}$, the isoform-specific curves collapsed onto a single curve (Figure 1F). We observed this collapse for the entire range of values of $N_{heads}$, $k_{on}$, and $k_{off}(0)$ given in Table 1 (data not shown). In all cases, the ensemble stall force $F_{max}$ matched the expected value given by:

$$F_{max} = F_{sm} N_{heads} dr(F_{sm}) \qquad (3)$$

where $F_{sm} = k_{x\text{-}bridge} d_{step}$ is the stall force for a single motor, and $dr(F_{sm}) = k_{on}[k_{on}+k_{off}(F_{sm})]^{-1}$ is the duty ratio of a single motor at stall. These data confirm that our model can reproduce isoform-specific variation in filament gliding speeds as well as the shape of force-velocity curves expected for all isoforms. Furthermore, they establish that for fixed values of crossbridge stiffness, step size, and force dependence of motor detachment, the behavior of simulated motor ensembles is determined by variation in three key quantities: the unloaded duty ratio $dr(0)$, the ensemble size $N_{heads}$, and the unloaded gliding speed $V_{max}$.

**The number of myosin heads, the motor duty ratio, and external force determine myosin filament processivity**

Single myosin II motors are non-processive, albeit to different degrees for different isoforms (20, 21, 23). Thus assembly into larger filaments is essential for stable engagement. We explored how varying $N_{heads}$ or $dr(0)$ affected the processivity of motor

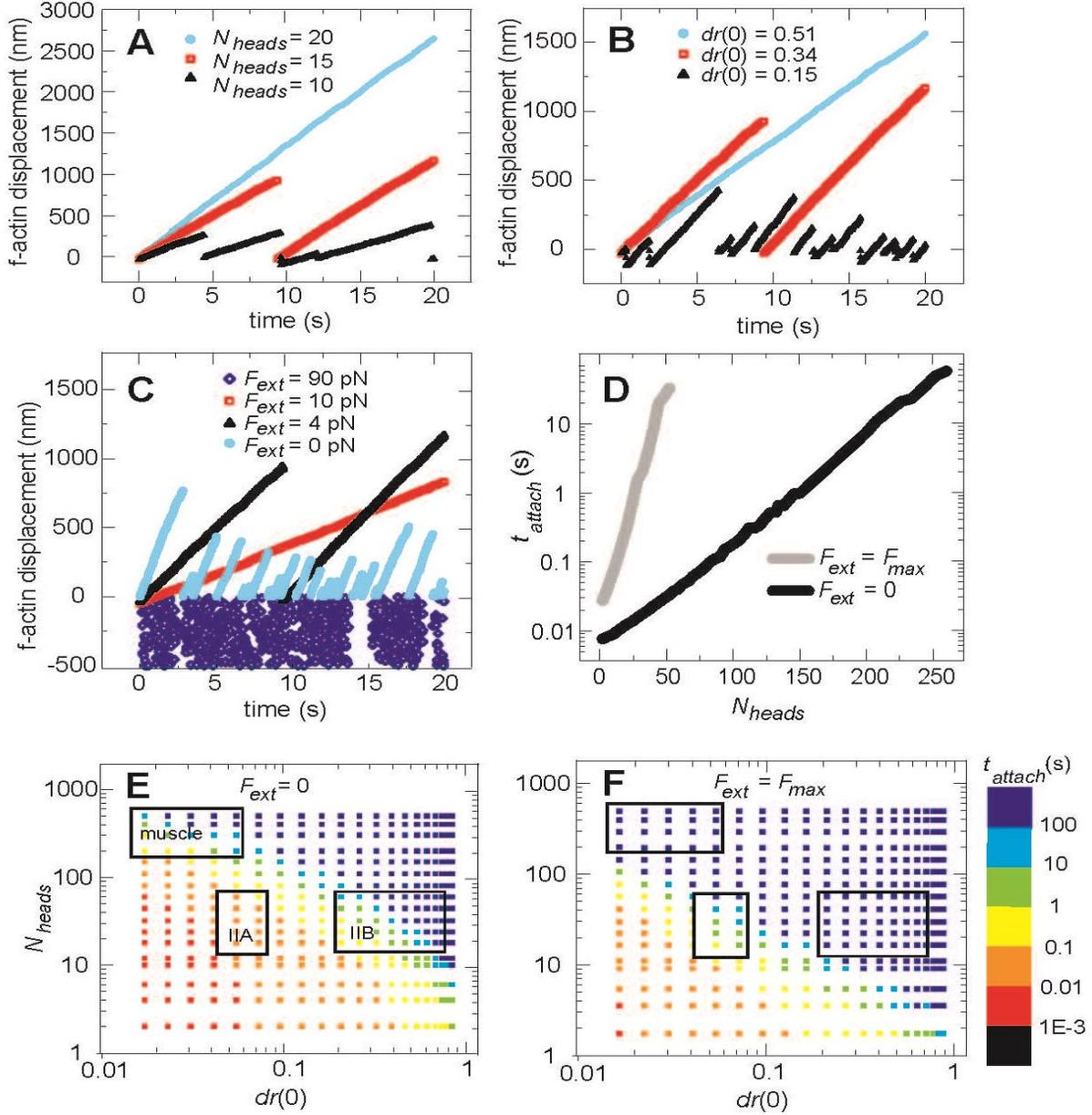

Figure 2: Dependence of myosin filament processivity on $N_{heads}$, $dr(0)$, and $F_{ext}$. (A) F-actin trajectories for varying values of $N_{heads}$. The F-actin was assumed to return to its original position upon release by the myosin. (B) F-actin displacements for varying $dr(0)$. (C) F-actin trajectories for varying $F_{ext}$. (D) The mean attached time as a function of $N_{heads}$ was shifted under stalled vs unloaded conditions. (E) Mean attached time on an unloaded actin filament for a range of $N_{heads}$ and $dr(0)$. The boxes indicate published values for muscle myosins, non-muscle myosin IIB, and non-muscle myosin IIA (see main text for references) starting at the top left and going clockwise. (F) Increased mean attached time for a stalled actin filament over same range as in (E). Parameter values in (A): $dr(0) = 0.05$ ($k_{on} = 10$ s$^{-1}$, $k_{off}(0) = 191$ s$^{-1}$), $F_{ext} = 4$ pN, (B): $k_{on} = 10$ s$^{-1}$, $N_{heads} = 15$, and $F_{ext} = 4$ pN, (C) $dr(0) = 0.34$ ($k_{on} = 10$ s$^{-1}$, $k_{off}(0) = 19$ s$^{-1}$), $N_{heads} = 15$, (D) $dr(0) = 0.05$ ($k_{on} = 10$ s$^{-1}$ and $k_{off}(0) = 191$ s$^{-1}$), (E) and (F) $k_{on} = 10$ s$^{-1}$

ensembles as measured by their mean attachment time to F-actin. As expected, varying either $dr(0)$ for fixed $N_{heads}$, or $N_{heads}$ for fixed $dr(0)$, could drive a transition from cycles of rapid attachment/detachment to processive translocation lasting more than 20 s (Fig. 2A-B). The mean attached time ($t_{attach}$) increased exponentially with $N_{heads}$ (Fig. 2D; black curve) or $dr(0)$ (not shown), consistent with previous results (38).

The force-dependence of ADP release should lead to an increase in the mean attached time in the presence of a resisting load (26-28). Consistent with this expectation, an increase the external load from 0 to 10 pN for fixed values of $N_{heads}$ and $dr(0)$ simultaneously slowed the motion of the F-actin and dramatically increased the attachment time (Fig. 2C). As anticipated from the form of equation (1), slip dominated at forces greater than the stall force of the myosin cluster and displacements of the F-actin were negative (Fig. 2C, diamonds).

To examine this effect further, we compared $t_{attach}$ as a function of $N_{heads}$ or $dr(0)$ for either unloaded filaments ($F_{ext} = 0$ pN) or filaments subjected to a stall force ($F_{ext} = F_{max}$) given by equation (3). The exponential rise in $t_{attach}$ with $N_{heads}$ (Fig. 2D) or $dr(0)$ (not shown), was significantly sharper for stalled versus unloaded filaments. This resulted in a dramatic increase in $t_{attach}$ going from unloaded to stalled conditions. The magnitude of this shift varied significantly with filament size and duty ratio as could be seen by measuring $t_{attach}$ across a range of $N_{heads}$ and $dr(0)$ for unloaded (Fig. 2E) and stalled filaments (Fig. 2F).

We identified regions of the phase spaces in Figs. 2E and 2F corresponding to experimentally measured ranges of filament size and motor duty ratio for skeletal or smooth muscle (14, 16, 23, 45), non-muscle IIA (12, 13, 20) and non-muscle IIB isoforms (12, 13, 21, 22). We found that the attachment times of the boxed regions could vary up to several orders of magnitude between the unstalled and stalled scenarios. This suggests that changes in myosin filament processivity due to forces from the surrounding environment could be significant for myosin II filaments *in vivo*.

**Motor properties and external stiffness shape the magnitude and stability of force generation**

From the analysis above, we expected the magnitude and stability of forces generated by myosin filaments in an F-actin network to depend on $N_{heads}$, $dr(0)$, and external forces. To examine this dependence, we considered a simple force-generating system consisting of one end of a bipolar myosin filament building force on an actin filament anchored by an elastic spring to its barbed end (Fig. 3A). We varied $N_{heads}$ or $dr(0)$ to tune filaments through the transition from weak attachment to processive engagement and monitored force buildup. For large $N_{heads}$ or $dr(0)$, the force increased steadily to approach the maximum stall force $F_{max}$, (Fig. 3B and 3C, blue squares). For intermediate values, the initial stages of force buildup were similar but both the maximum force and the stability of force buildup was compromised (Fig. 3B and 3C, red circles). Rather than generating stable tension on the F-actin, the filament detached

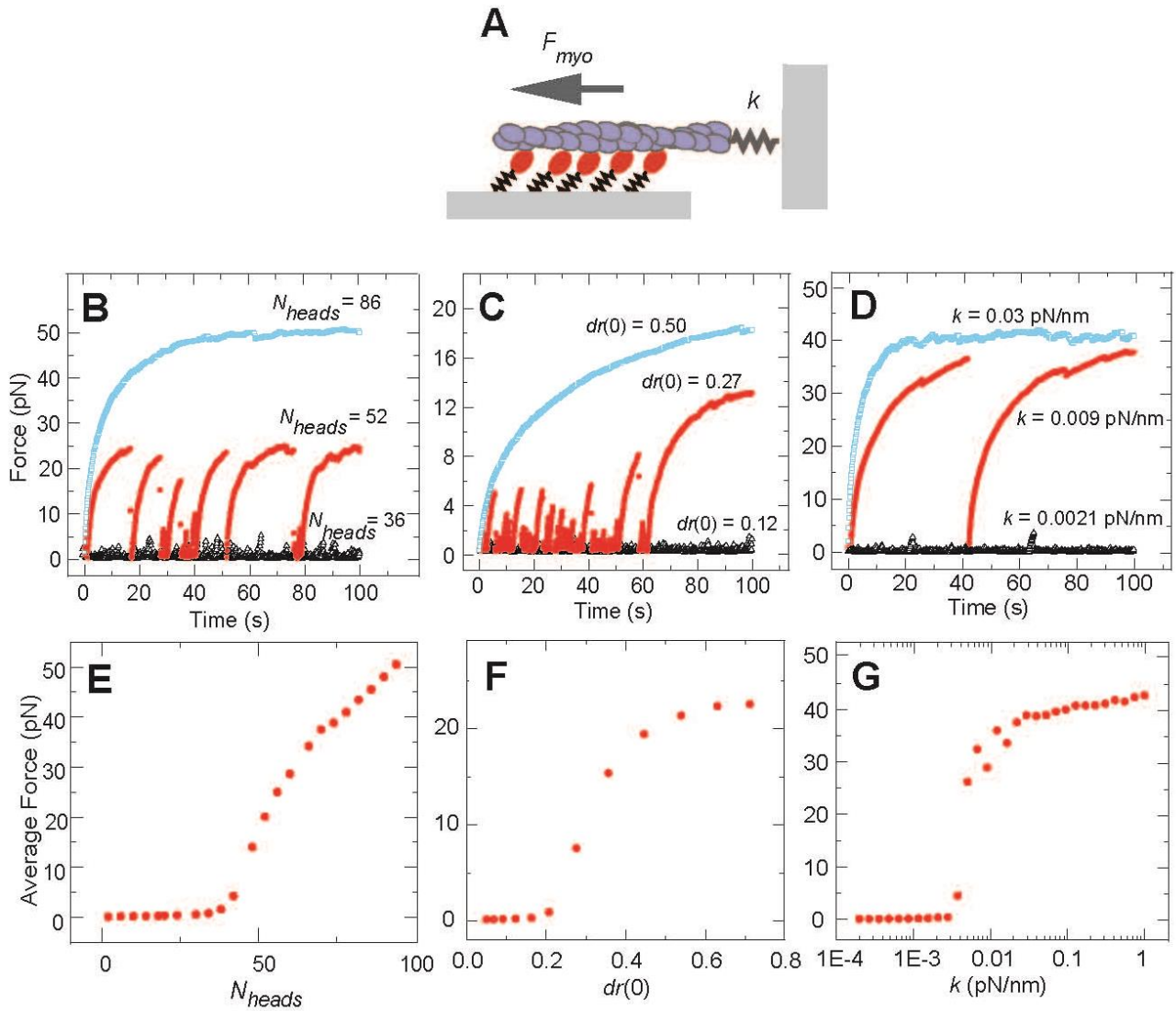

Figure 3: Myosin filament size, motor duty ratio, and actin stiffness determine the magnitude and stability of force generation. (A) Simulations were arranged with fixed myosin heads building force on F-actin anchored to a spring. (B) Increasing the size of the myosin filament produced a transition from transient force buildup and release to stable force maintenance. (C) Similarly, the processivity of force generation increased with the unloaded duty ratio of the motors. (D) Increasing k also increased the stability of generated force. (E) The average force on the spring in ten 100 s simulations for different values of $N_{heads}$. (F) and (G) A sharp increase of the average force was also produced with increasing $dr(0)$ and k. Parameter values in (B) and (E): $dr(0) = 0.05$ ($k_{on} = 10$ s$^{-1}$, $k_{off}(0)=191$ s$^{-1}$) $k = 0.02$ pN/nm, (C) and (F): $k_{on} = 10$ s$^{-1}$, $N_{heads} = 10$, $k = 0.02$ pN/nm, (D) and (G): $dr(0) = 0.05$ ($k_{on} = 10$ s$^{-1}$, $k_{off}(0) = 191$ s$^{-1}$), $N_{heads} = 74$

repeatedly such that force buildup alternated with intermittent periods of release. At the smallest values of $N_{heads}$ or $dr(0)$, force buildup was completely lost (Fig. 3B and 3C, open black triangles).

Plotting average force against $N_{heads}$ or $dr(0)$ revealed dramatic differences in the average force generated by myosin filaments. When either $N_{heads}$ or $dr(0)$ was sufficiently low to abrogate processive runs, the average force remained approximately zero (Fig. 3E and 3F). Above a threshold value, the average force increased rapidly with $N_{heads}$ or $dr(0)$ to approach the myosin filament stall force $F_{max}$ (Eqn (3)). Changes to either parameter thus produced a transition from a state where small attachment times were insufficient to generate considerable force to one in which the myosin remained attached long enough to build to stall.

Finally, we explored how variation in spring stiffness modulates force buildup. Figure 3D illustrates the response to varying stiffness of a myosin filament with constant $N_{heads}$ and $dr(0)$. Similar to the effect seen with decreasing $N_{heads}$ and $dr(0)$, reducing $k$ produced a transition from stable force generation to intermittent force build up and release and finally to complete inhibition of force generation. Plotting average force vs. stiffness revealed a sharp transition from ~ 0 to maximal force for a ~4-fold change in $k$ (Fig. 3G). We observed similarly sharp transitions for other choices of motor parameters, although the value of $k$ at which transition occurs depends on both $N_{heads}$ and $dr(0)$. Qualitatively, this effect can be understood as a competition between the timescale of myosin processivity and the rate of force buildup set by spring stiffness. The sharp increase in force output occurs when $t_{attach}$ exceeds the time required to build to stall.

**Determinants of characteristic time scale of force buildup**

To further understand the regulation of force buildup shown in Fig. 3, we explored how quickly the maximum force was built within the processive regime. We measured the time to build 70% of the maximum force ($t_{build}$) as a function of $F_{max}$, $V_{max}$ and $k$. As expected, faster motors built force more rapidly such that $t_{build}$ scaled linearly with $1/V_{max}$ (Fig. 4A). Alternately, for a constant motor speed, $t_{build}$ should increase in proportion to the number of steps required to reach the stall force; consistent with this, $t_{build}$ was directly proportional to $F_{max}$ (Fig. 4A) and inversely proportional to k (Fig. 4B). Indeed, we observed a single scaling relationship for $t_{build}$ as a function of all three parameters (Fig.4C):

$$t_{build} \sim \frac{F_{max}}{kV_{max}} \quad (4)$$

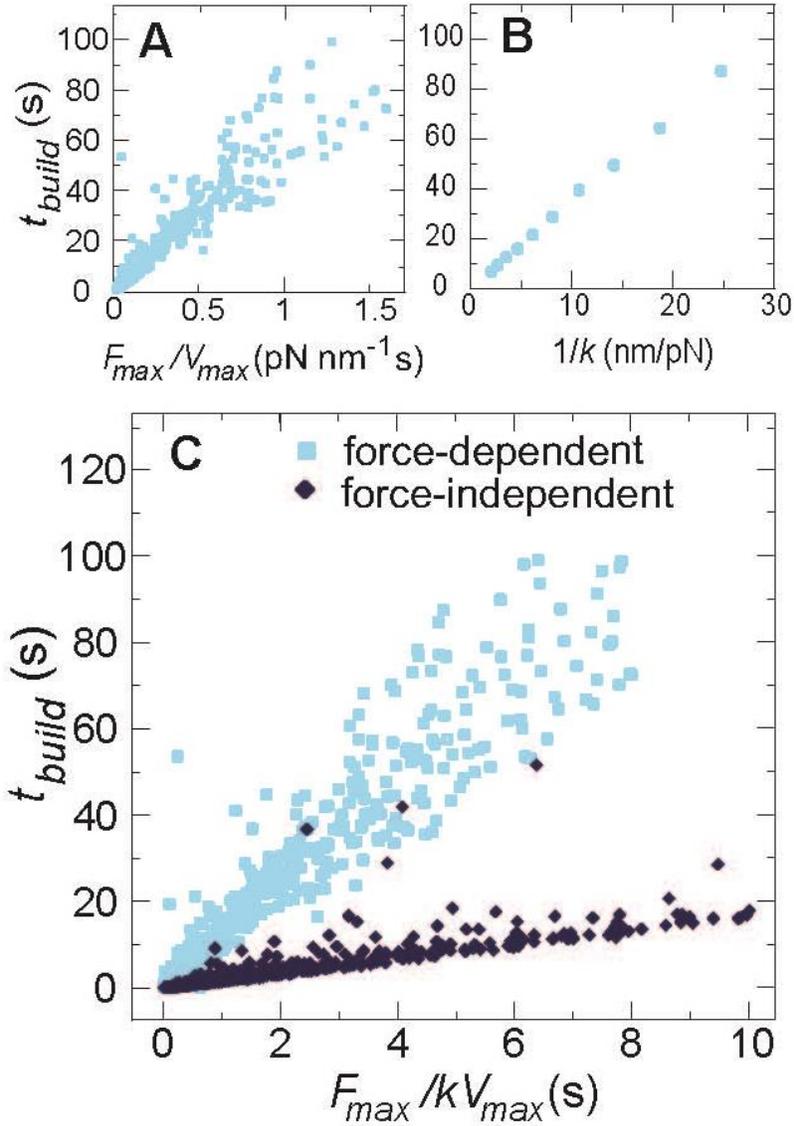

Figure 4: The characteristic time for processive motors to build force ($t_{build}$) scales as $F_{max}V_{max}^{-1}k^{-1}$. (A) A linear increase of $t_{build}$ was observed with $F_{max}/V_{max}$ with $k = 0.01$ pN/nm. The values of $k_{off}(0)$, $k_{on}$, and $N_{heads}$ were varied from 10 s$^{-1}$ to 191 s$^{-1}$, 1 s$^{-1}$ to 10 s$^{-1}$, and 6 to 600 respectively. (B) $t_{build}$ also increased linearly with $1/k$. Parameter values: $N_{heads} = 50$, $dr(0) = 0.05$ ($k_{on} = 10$ s$^{-1}$ and $k_{off}(0) = 191$ s$^{-1}$). (C) The final scaling relationship for both force-dependent and force-independent motors. The same ranges of $k_{off}(0)$, $k_{on}$, and $N_{heads}$ from (A) were used with $k$ ranging from 0.001 to 1 pN/nm.

Interestingly, we observed a similar scaling when we removed the force dependence of myosin release, albeit with a lower slope (Fig. 4C, dark blue diamonds). Thus myosin motors build force more quickly on stiffer substrates regardless of the exact mechanochemistry. This suggests that force-dependent kinetics may not be required for experimentally observed increases in the rate of force generation with external stiffness of contractile cells as has been previously assumed (11, 46, 47).

**Force-dependent myosin kinetics produced a switch-like transition from non-processive to processive force generation**

The analysis shown in Fig. 3E and 3F revealed a very sharp increase in the average force with increasing $N_{heads}$ or $dr(0)$. We sought to understand how the sharpness of this transition depends on mechanosensitivity of the myosin duty cycle. For myosin filaments building force against an elastic load, the mean attachment time ($t_{attach}$) should fall between the two extreme values measured under unloaded or stalled conditions (Fig. 2D). Indeed, $t_{attach}$ for relatively small myosin filaments (Fig. 5A, open squares) resembled that of the unloaded motors (Fig. 5A, black line). Increasing $N_{heads}$ produced a sharp, faster-than-exponential increase in $t_{attach}$ (Fig. 5A, open squares) that coincided with a sharp increase in average force (Fig. 5B, open squares). Absent force-dependent kinetics, both the faster-than-exponential increase in $t_{attach}$ and the sharp increase in force were completely abolished and the dependence of mean attachment times on $N_{heads}$ was very similar to that of unloaded motor ensembles (compare red diamonds vs. dark blue open squares in Fig. 5A, B). Thus, the number of motors required to generate a given level of force was significantly higher for motors lacking force-dependent kinetics and the rate of force increase with $N_{heads}$ at a threshold value was lower.

These data reveal how force-dependent kinetics mediate a form of positive feedback in which force buildup promotes increased attachment and further force-buildup. This feedback sharpens the effect of increasing duty ratio or filament size such that small increases in either quantity above threshold values causes a rapid transition from a state in which transient attachments produced little force to one in which force was built and maintained over long timescales. As a consequence of this feedback, large force fluctuations depicted by the red circles in Fig. 3B and 3C occur only within narrow ranges of $N_{heads}$ or the motor duty ratio.

**Mechanical cues regulate the switch to processive force generation**

Another potential consequence of this feedback is increased sensitivity of force production to environmental stiffness or forces. As already shown (Fig. 3G), increasing external stiffness $k$ can drive a sharp transition from non-processive actomyosin interactions to processive force buildup. The sharp transition is completely abolished in motor filaments lacking force-dependent kinetics (red vs. dark blue traces in Fig. 5D). The sharp increase in force production with increasing stiffness can be understood in terms of the interplay between network stiffness and force-buildup times characterized in Figure 4: motor filaments build force faster on stiffer substrates and thus engage positive feedback.

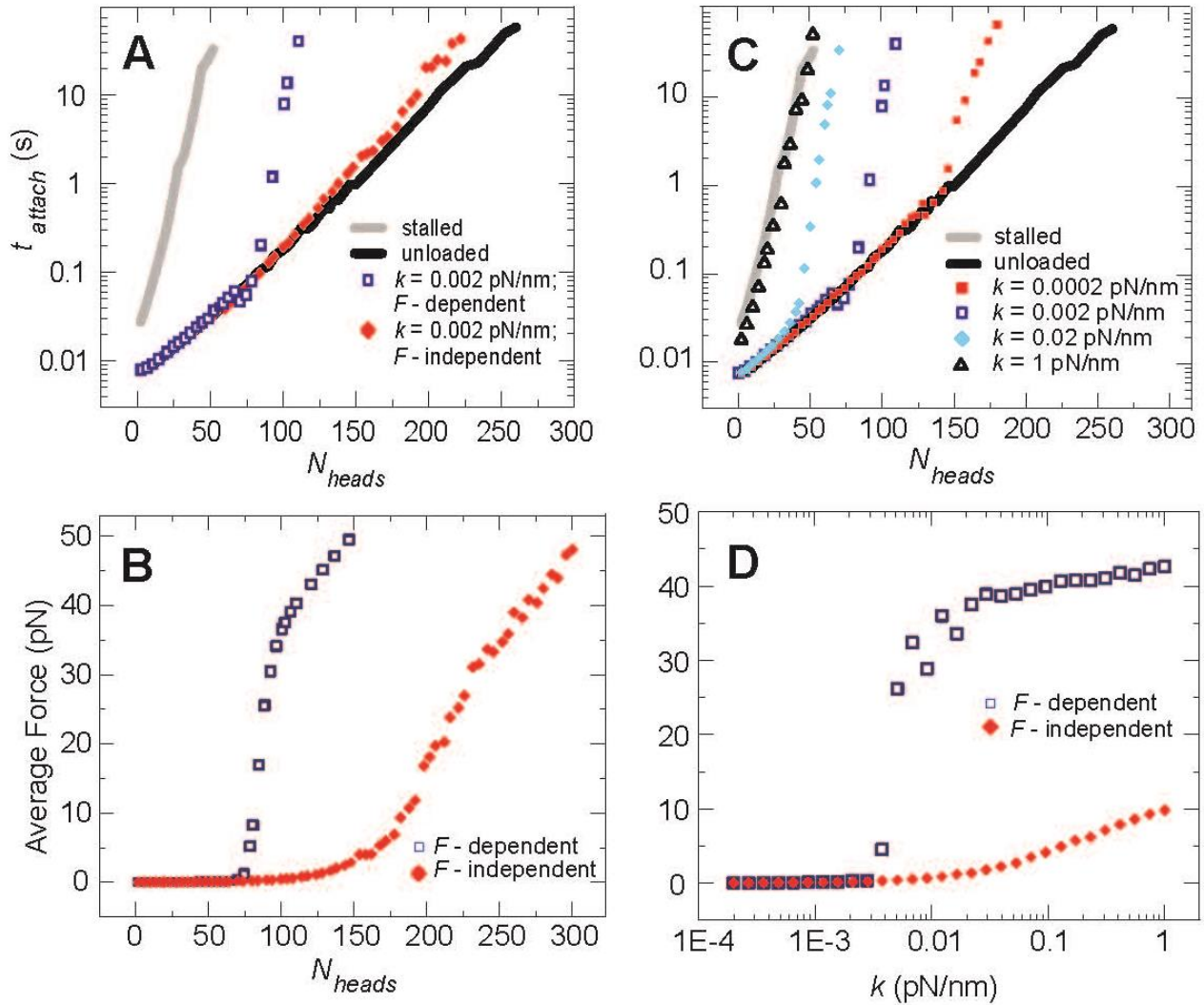

Figure 5: Force-dependent myosin kinetics produce switch-like transition into processive state. (A) Increase of the average attached interval, $t_{attach}$, with $N_{heads}$ for unloaded, stalled, and spring-loaded actin filaments. For the spring-loaded case, curves in which force-dependence of $k_{off}$ was either included or neglected are shown. Averages were taken over fifteen 200 s simulations. (B) Average force output in the spring-loaded cases from (A). (C) The steep increase of $t_{attach}$ for spring-loaded actin filaments was shifted by varying $k$. (D) Force production at a single value of $N_{heads}$ was sharply dependent on stiffness when $k_{off}$ was force-dependent. Force-independent motors showed a weaker dependence on $k$. $N_{heads}$ = 50. In all panels, $dr(0) = 0.05$ ($k_{on} = 10$ s$^{-1}$ and $k_{off}(0) = 191$ s$^{-1}$).

We found that different levels of external stiffness can shift the threshold for engaging positive feedback with increasing filament size (Fig. 5C) or duty ratio (not shown). In very stiff environments (i.e when $k$ exceeds the myosin cross-bridge stiffness of 0.7 pN/nm), $t_{attach}$ was similar to that expected from stalled motors (Fig. 5C, gray line) because the motors reached stall very quickly (Fig. 5C, open triangles). In softer environments, the threshold filament size required to engage positive feedback increased with decreasing $k$ from $N_{heads}$ ~1 for $k = 1$pN/nm to $N_{heads}$ ~40 for $k = 0.02$ pN/nm to $N_{heads}$ ~150 for $k = 0.0002$ pN/nm (Fig. 5C).

Finally, we explored the extent to which positive feedback due to force-dependent kinetics renders force production sensitive to externally applied forces transmitted through an elastic network. We held $k$ constant and applied a small constant load ($F_{ext}$) to the F-actin (Fig. 6A). As shown in Fig. 6B, increasing $F_{ext}$ from 0% to 7% of the myosin filament stall force $F_{max}$ reduced the threshold filament size required to engage positive feedback and transition from non-processive to processive engagement from $N_{heads}$ ~100 to $N_{heads}$ ~60. Alternatively, increasing the externally applied force for fixed motor parameters and filament size produced a very sharp increase in average force over a narrow range of applied forces. For the motor parameters used in Fig. 6C, an increase in $F_{ext}$ from 0 to 1 pN (about 5% of the stall force) produced an increase in the average force from ~0 to $F_{max}$. Thus, positive feedback due to force-dependent myosin kinetics make force production on an elastic substrate highly responsive to relatively small variations in applied force.

**Myosin II isoform performance in elastic networks**

To explore the potential consequences of the behaviors described in Figs. 2-5 on different myosin II isoforms, we utilized parameters that reflected $N_{heads}$ and enzymatic rates of skeletal muscle, nonmuscle IIA, and nonmuscle IIB and considered the timescales of attachment and force buildup as the environmental stiffness $k$ was varied.

For parameters consistent with skeletal muscle myosin filaments (Table 2, Fig. 1), the unloaded $t_{attach}$ was approximately 70 s. As $k$ varied from 0.001 to 1 pN/nm, $t_{build}$ decreased from ≈ 100 s to 0.1 s, while $t_{attach}$ rapidly increased to values exceeding our simulation time of 1000 s (Fig 7A). Thus, the combination of the large filament size and high speed of skeletal muscle myosin II allowed for rapid and stable build up of force over a wide range of stiffness.

Using motor parameters and filament size appropriate for non-muscle myosin IIB yielded stable attachment ($t_{attach}$>1000 s) in unloaded conditions for all values of $k$ (Fig. 7C). However, $t_{build}$ also was >1000 s for $k < 0.1$ pN/nm, decreasing to 100 s only for $k > 1$ pN/nm. These data suggest that myosin IIB is well-tuned to function as a high-affinity cross-linker over a wide range of environmental stiffness and force, as has been speculated previously (20, 21). However the exceedingly slow force buildup time suggests that myosin IIB will be very ineffective at generating force against actin networks that turn over on timescales relevant for rapid morphogenic change (1-100 s; see discussion).

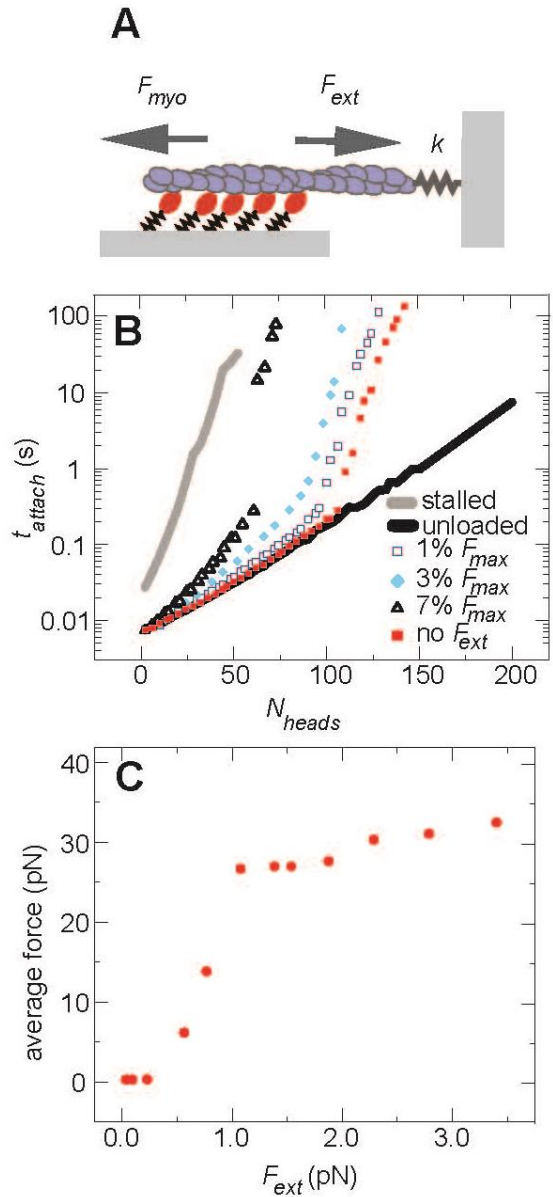

Figure 6: Small variations in a constantly applied force can trigger processive force generation. (A) Motors build force on an elastic load in the presence of an additional constand load, $F_{ext}$ (B) The average attached time of myosin filaments to spring loaded F-actin with varying $F_{ext}$. (C) Increase of the myosin generated force with $F_{ext}$. Parameter values in (B): $dr(0) = 0.05$ ($k_{on} = 10$ s$^{-1}$ and $k_{off}(0) = 191$ s$^{-1}$), $k = 0.0006$ pN/nm (B): $dr(0) = 0.05$ ($k_{on} = 10$ s$^{-1}$ and $k_{off}(0) = 191$ s$^{-1}$), $k = 0.0006$ pN/nm, $N_{heads} = 100$.

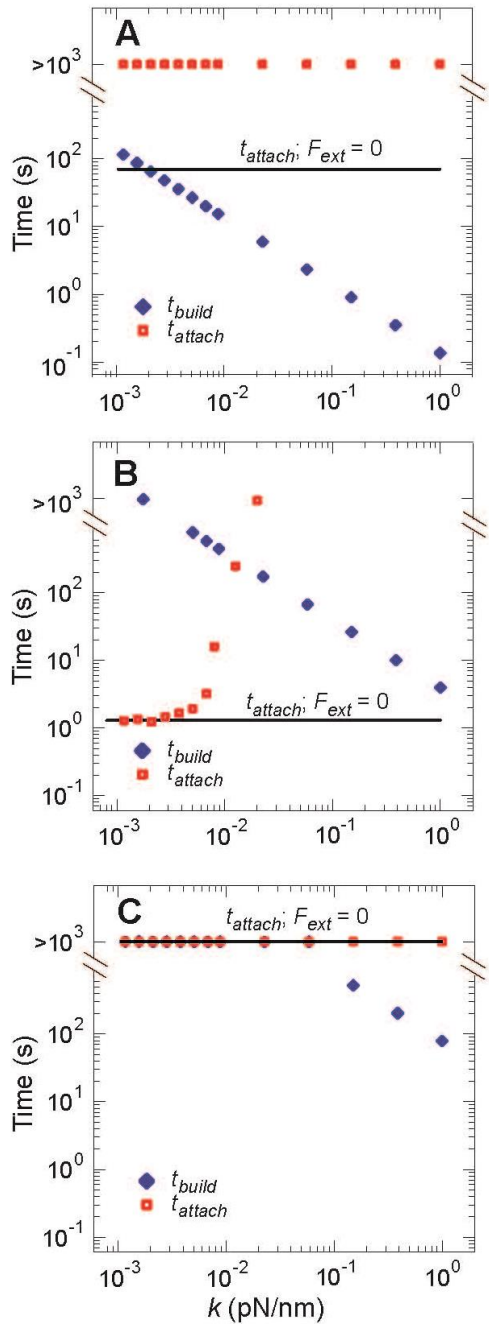

Figure 7: Variations in attachment and force buildup times for different myosin isoforms yield qualitatively distinct behaviors in an elastic environment. (A) For motor parameters chosen to match skeletal muscle myosin, $t_{attach} >= t_{build}$ for all values of $k$. (B) For motor parameters chosen to match non-muscle myosin IIA, a sharp transition from transient attachment and minimal force buildup to stable force generation above a threshold $k \sim 0.01$ pN/nm occurred when $t_{attach}$ becomes long enough to engage positive feedback (see text for discussion). (C) Parameters chosen to match non-muscle myosin IIB type motor yielded much longer time scales for both $t_{attach}$ and $t_{build}$. For all panels, the default parameters in Table 1 and isoform-specific parameters from Table 2 were used. Values of $N_{heads}$ in (A): 500, (B): 50, (C): 50.

Intriguingly, when we chose parameters appropriate for nonmuscle myosin IIA filaments, our simulations predicted qualitatively distinct behaviors at low and high $k$ (Fig. 7B). Myosin IIA filaments are predicted to bind processively at stall. However, for $k < 0.01$ pN/nm, the time required to build force was too long to engage positive feedback and switch to stable attachment, so the attachment time remained quite short ($t_{attach} < 1$ s). Around 0.01 pN/nm, a sharp transition to processive force build up occurred as $t_{build}$ decreased from 100 to 1 s, with a concomitant increase in $t_{attach}$. Thus, parameters consistent with myosin IIA suggest that it may be poised to serve as a low affinity cross-linker at low stiffness but a processive force generator at high stiffness.

**Discussion**

Cross-bridge models have been used extensively to model force generation by skeletal muscle contracting against a constant load (29, 32-37). Here, we extended this approach to explore force generation by myosin filaments on an elastic substrate. In particular, we have explored how the production and maintenance of force is influenced by the interplay of filament size, force-dependent mechanochemistry, and substrate stiffness. We first verified our model using two classic tests of myosin performance: the unloaded gliding assay and their force-velocity relationship. We were able to obtain gliding velocities consistent with those found experimentally for skeletal muscle, smooth muscle, and non-muscle isoforms by using experimentally determined binding and unloaded unbinding rates while maintaining constant parameters for the cross-bridge stiffness, step size, and force dependence. The force velocity curve displayed the appropriate concave shape expected from studies of muscle fibers (44) and motor clusters *in vitro* (43). Both the unloaded velocities and force-velocity relationships fell onto master curves when scaled appropriately by three key properties: $V_{max}$, $N_{heads}$, and $dr(0)$. We went on to show that these properties, in addition to the force-dependent unbinding rate, were critical in determining the motors' ability to produce and maintain tension and sense mechanical cues in their environment.

We compared the mean attached time of motor clusters under unloaded and stalled conditions to assay their sensitivity to local mechanics. We found an increase in $t_{attach}$ for stalled versus unloaded motor clusters. The magnitude of this shift increased rapidly with $N_{heads}$ and $dr(0)$. For ranges of $N_{heads}$ and $dr(0)$ corresponding to different myosin II isoforms, this difference in $t_{attach}$ was up to several orders of magnitude. Thus, sharp sensitivity to force is a unique property of clusters of non-processive motors compared to the processive dimers formed by other myosin isoforms.

The mean force output of myosin motors in a network will generally depend on how quickly they build force compared to timescales associated with relaxation events including their own detachment, actin cross-linker unbinding, and actin turnover. We found the characteristic time of building to stall on an elastic load with stiffness $k$ to be described by a simple relationship $t_{build} \sim F_{max}/k*V_{max}$ that holds for both processive and nonprocessive motors. Tuning of either $t_{build}$ or relaxation times provides a means to regulate whether a given isoform will generate force quickly enough to be effective for contractility. Examples of potential regulatory mechanisms include the nonlinear stiffness of actin networks, which is highly dependent on the level of internal or external pre-stress and the network connectivity (48). Force relaxation likely depends on both cross-link unbinding (49, 50) and F-actin turnover (51) and may occur on the order of 1-100 s.

While the regulation of force production via myosin filament properties or network mechanics does not require force-dependent kinetics, these force-dependent kinetics enable coupling between force production and relaxation by myosin filament unbinding. Our simulations predict that at threshold values of $N_{heads}$, $dr(0)$, or $k$, positive feedback between force generated on the elastic environment and motor attachment produced switch-like activation of processive force production. We expect that there are many ways in which cells could tune myosin filaments into a regime where small forces engage this feedback and effectively turn on contractility. The size and density of myosin filaments, affinity of myosin for different actin network geometries, and actin network viscoelasticity may all vary significantly and be regulated spatiotemporally. The presence of myosin filaments at a sufficiently high density could allow individual filaments to transmit force to one another and cooperatively switch into a processive state. Additionally, a myosin filament in a heterogeneous environment could selectively bind and build force on stiffer or more stable actin structures due to the feedback conferring increased affinity.

Our results allowed us to predict the types of behaviors displayed by different myosin II isoforms in physiological environments with varying stiffness. As expected, motor clusters with the high value of $N_{heads}$, high $V_{max}$, and low $dr(0)$ characteristic of skeletal muscle myosin processively built force over the entire range of stiffness we explored. In contrast, we expect the nonmuscle myosin isoforms to show greater selectivity for stiffness or other mechanical signals for different reasons. In the case of a nonmuscle IIA filament with $N_{heads} = 50$, the attachment time showed a steep increase in attachment time and average force with increasing stiffness from a small unloaded value to maximum (stalled) force, consistent with the positive feedback described above. For myosin IIB, the attachment time even for an unloaded filament was over 1000 s. However, force generation on soft substrates may still be limited due to the long time scale of force build up, and in this regime the filaments may primarily function as actin cross-linkers. While it has been proposed that the non-muscle myosin IIA and non-muscle myosin IIB duty cycles were better tuned for tension generation and tension maintenance respectively (20, 21), to our knowledge this is the first example demonstrating this behavior with experimentally measured parameters and revealing its dependence on stiffness. We note that these results depend in a nonlinear fashion on the exact value of $N_{heads}$ used and modest changes can tune the behavior of an isoform into another regime. For example, reducing the value of $N_{heads}$ to 200 using the skeletal muscle myosin parameters restored the switch-like transition to high processivity from relatively small, unloaded values (data not shown).

Finally, our results have implications for coarse-graining of myosin activity in simulations and analytical work. Alternative representations of myosin activity as either time-independent force dipoles (52) or force dipoles that transiently pull and release (53) within a continuum elastic or fluid medium have been used for applications such as predicting the strain field from interacting dipoles and mechanical properties of active networks. Our results suggest that the appropriate representation will depend on the myosin isoform and the mechanical context in which the motor operates. More detailed representations that allow the dipole kicking rate to depend on force may be essential to capture force-dependent dynamics that underlie large-scale deformations of an actomyosin network (54). How motor properties influence an actomyosin network's ability to produce force or change its shape, or how they may modulate such activity if the network is subjected to external force or tethering to an external substrate are interesting questions for future study.

**Acknowledgments**
S.S. was supported by the National Institute of Biomedical Imaging and Bioengineering of the National Institutes of Health under Award Number T32EB009412.  M.L.G. is supported by the Packard Foundation and American Asthma Foundation Early Excellence Award. This work was supported by the U. Chicago MRSEC.